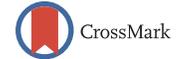

# Research Article
# Control and Data Flow Execution of Java Program

Safeeullah Soomro, Zainab Alansari and Mohammad Riyaz Belgaum

College of Computer Studies, AMA International University, Building 829, Road 1213, Block 712, P.O. Box 18041, Salmabad, Kingdom of Bahrain

## Abstract

**Background and Objective:** Since decades understanding of programs has become a compulsory task for the students as well as for others who are involved in the process of developing software and providing solutions to open problems. In that aspect showing the problem in a pictorial presentation in a best manner is a key advantage to better understand it. **Materials and Methods:** This article provides model and structure for Java programs to understand the control and data flow analysis of execution. Especially it helps to understand the static analysis of Java programs, which is an uttermost important phase for software maintenance. This article provided information and model for visualization of Java programs that may help better understanding of programs for a learning and analysis purpose. The idea provided for building visualization tool is extracting data and control analysis from execution of Java programs. **Results:** Theoretically, this article has shown how to extract source code and provide information of data and control graph. It is helpful for understanding of programs and may help towards software debugging and maintenance process. **Conclusions:** This article presented case studies to prove that our idea is most important for better understanding of Java programs which may help towards static analysis, software debugging and software maintenance.

## INTRODUCTION

The Understanding of Programs is most importance factor of computer science and software engineering students as well as the developers who are solving open questions. Since decades' people were working on program analysis, software maintenance and software re-engineering to provide some intelligent tools which were beneficial for everyone. But still lack of tools and techniques which may provide better understanding of program visually or pictorial presentation, which may provide ease to learn programming to students of current era as well as for the future. A lot of research has been conducted in this regard but still needs more time and effort to overcome problems for better understanding of problems or generally program analysis or software maintenance. Also, it is extremely key area of research in the field of computer science and software engineering because manual understanding of programs is very difficult to understand and time consuming. It is cost effective and most part of projects funding is involved since last decade. In this paper, technique which may provide execution of Java program's information in data, program and control dependencies are presented, which may help to understand the programs and reduce time and cost for software maintenance and debugging in future.

Program analysis is the process of automatically understanding and presenting real behavior of programs which is most important for research community towards software analysis and maintenance.

Singer and Kirkham[1] and Du *et al*.[2] provided program information in dependencies and visualize the programs. Luijten[3] has presented a viable visualization tool, Coffee Dregs, for Object oriented programs that supports multi-threaded Java programs, with standard input and output and GUI-programs to a limited extent. Some of the research work has been done[4,5] in path conditions and analysis techniques, they have provided model for the understanding of programs in static way but not provided application.

Huizing *et al*.[6] and Gestwicki and Jayaraman[7] provided model for object oriented programs together with visualization but it is a tool which shows relationship between user and machine. It is a special tool for e-learning purpose which may help others to understand graphical statements. Jayaraman and Baltus[8] provides execution according to programs paths and it is done dynamically making the best way of code executed visually. Our idea is to visualize programs in terms of data, control and program graphs which option is not available in both tools[8,9] according to our knowledge.

Future research has to apply this theoretical concept into practical approach to develop an automatic and intelligent tool, which is really today's need in the world. This article has presented a program analysis of Java programs in terms of execution. This article has extracted source code from Java programs and found path execution of programs in terms of data and control dependences. After that This article have visualized source code into blocks and nodes so that provides information flow graphs, allowing the users and learners of Java programs a better understanding.

## MATERIALS AND METHOD

**Control and data flow analysis:** Many tools have been provided for program analysis and software maintenance but still lacking to understand the programs particularly in regard of control and data. That means static analysis of programs which is useful for this area of research where an automatic and intelligent tool can be made. Also provided is the exact information of the execution of programs to understand the data and control flow of source code. There are some tools provided but about our idea there is no any tool which may provide better understanding of programs in terms of data and control flow of programs. The idea is to reduce the cost of Software Analyst and Coders.

This article provided better model and technique which may affect better understanding of programs with regards of data and control graphs particularly, which can be an addition towards research community for better understanding of code in visuals or pictorial presentation making it the most important for learning process in all fields of science and technology.

In the Fig. 1 this article has presented the steps towards our idea and do work accordingly. The First step is to extract source code using AST (Abstract Syntax Tree). Afterwards information extracted from path execution of programs in static forms. This article provides an algorithm to make branches and nodes accordingly, then generated control graph of the whole path execution in visual form.

To understand the code is essential for testing, reverse-engineering and maintenance. It is also useful for the deeply understanding of programs which leads to produce an error free program. This is also true that researchers have carried research towards analysis and maintenance of programs. Since decades many models and techniques are available for the researchers[6-8] but still lacking in particular tools for visualizing programs in terms of understanding the control flow, data flow and program flow according to source code. Although it's very time consuming task but its today's demand of the market for providing such tools which can lead to accurate analysis of software. To overcome this problem, visualization tool for Java programs should be designed, implemented and tested.

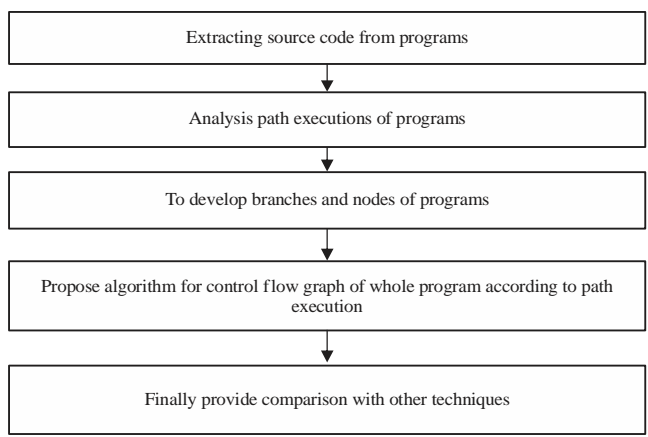

Fig. 1: Steps of the model for Java programs visualization

The Goal of this work is to provide better technique to all researchers in the world to carry source code for the analysis of control and data flow of Java programs. In Fig. 2, it provides the model structure of our tool showing how to deal with it and provide better solution for visualization of programs. It is presented that how to visualize the source code and extract its Abstract Syntax Tree (AST). After that code is converted into graphs (Data, Variable and Program). Our method is to extract information from AST and calculate the dependencies of variables like Line 1: a = b, Line 2: b = c, Line 3: c = a+b, count as Branch B1, B2 and B3 and Node N1. After extracting code from AST then it may provide nodes and branches of all program statements according to program execution so that it cannot ignore any program statement which may affect the source code of the program.

Finally, this study provides technique or method which can build plug-in tool in future to add on eclipse for the Java and provide data, control and program dependence graphs for better understanding of programs. This presented technique may help academia, industry and others to get benefit for the software analysis, maintenance and debugging as well in future.

## RESULTS

This section contains information of the program execution and representation in control ow graph and data dependence graph from the source code. This article presented analysis according to path executions of program regarding the data and control. There are two kinds of the execution of programs named as static and dynamic.

Static execution provides whole text of the program for analysis. Static always provides all information of program having all control ow possibility according to the source code. Java program were extracted all possible path executions. In the example program, it found four path executions according to true and false values for those conditions. This article has shown the static path executions of our program 1 as under:

Line Number 3: x > y: FALSE
Line Number 7: y > 5: FALSE
Execution Path 1: 0 1 2 3 5 6 7 9 10
Line Number 3: x > y: TRUE
Line Number 7: y > 5: FALSE
Execution Path 2: 0 1 2 3 4 6 7 9 10
Line Number 3: x > y: FALSE
Line Number 7: y > 5:
TRUE Execution Path 3: 0 1 2 3 5 6 7 8 10
Line Number 3: x > y: TRUE
Line Number 7: y > 5: TRUE
Execution Path 4: 0 1 2 3 4 6 7 8 10

Dynamic execution provides the exact ow control of program according to source code of program execution. It depends on compiler to compute and execute program statements based on the input values and other control ow statements of the program. This article has presented dynamic execution path of our program 1 as under:

Line Number 3: x > y: FALSE
Line Number 7: y > 5: FALSE
Execution Path: 0 1 2 3 5 6 7 9 10

In the Fig. 3 an example of Java program is written and This article have shown the execution passing through all paths. Our approach is to derive control ow graph and

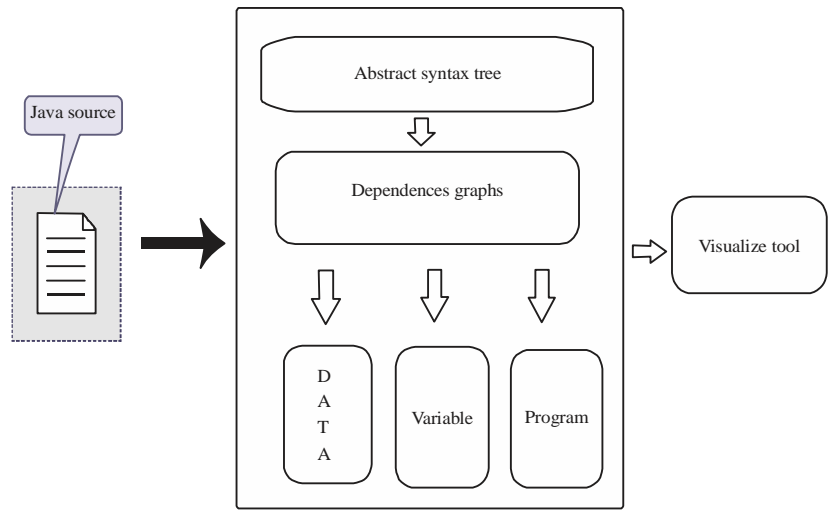

Fig. 2: Block diagram of the model for Java programs visualization

```
▷ Public class Test Program {
public static void main (string[]args) {
     0 : int x  =  3;
     0 : int x  =  3;
     1 : int y  =  4;
     2 : int z  =  0;
     3 : if (x > y);
     4 : z = y + 2;
     5 : else z = y + 2;
     6 : z = z + y;
     7 : if  (y > 5)
     8 : z = z + 5;
     9 : else z = z – 2;
     10 : z = z + 3;
         } }
```

Fig. 3: Simple Java example

dependence graph from its source code. All statements were extracted from the source code and have made blocks and nodes of all statements. The program has been compiled and analysed source code.

**DISCUSSION**

Java tool[9,10] called as Insight which presented runtime behaviour of programs and helps for debugging purpose. However, it helps the execution of Java programs dynamically but cannot provide information statically which may help research community towards software analysis and maintenance.

This study suggests tool which may provide program execution in terms of program, data and variable dependencies in visual forms. This article assumed to present visualization of all statements of Java programming language like method calling, object creation, calling object, parameters passing through methods and objects, polymorphism and others in object oriented program. visual presentation in this study provides full details for the basic lines of code, multiple lines, loops, nested structures of code and fully support to object oriented programs.

This article has discussed our approach and provided an example regarding execution so that it may be easy to extract data and control flow of programs accordingly. Theoretically, this article has shown how to extract source code and provide information of data and control graph using a simple example. In future, tool can be developed for more results analysis and discussion which may be helpful for the research community and as well as students to learn Java program in an efficient way.

**CONCLUSION**

This article has presented data and control information of Java programs using static and dynamic execution. It is helpful for understanding of programs and may help towards software debugging and maintenance process. Moreover, this study presented a theoretical idea towards visualization of the source code information of programs which is an essential for students as well as developers to understand the programs. Presented case studies may help readers towards enhanced better understanding capabilities of programs.

**SIGNIFICANT STATEMENT**

This research work discovered that the idea provided for building visualization tool is extracting data and control analysis from execution of Java programs.